\documentclass[sigconf]{acmart}
\AtBeginDocument{%
  \providecommand\BibTeX{{%
    \normalfont B\kern-0.5em{\scshape i\kern-0.25em b}\kern-0.8em\TeX}}}

\copyrightyear{2022}
\acmYear{2022}
\setcopyright{acmlicensed}\acmConference[CHI '22 Extended Abstracts]{CHI Conference on Human Factors in Computing Systems Extended Abstracts}{April 29-May 5, 2022}{New Orleans, LA, USA}
\acmBooktitle{CHI Conference on Human Factors in Computing Systems Extended Abstracts (CHI '22 Extended Abstracts), April 29-May 5, 2022, New Orleans, LA, USA}
\acmPrice{15.00}
\acmDOI{10.1145/3491101.3519858}
\acmISBN{978-1-4503-9156-6/22/04}

\begin{document}

%%
%% The "title" command has an optional parameter,
%% allowing the author to define a "short title" to be used in page headers.
\title[Investigating Potential of AI to Support Different Memory Types for PwD]{Investigating the Potential of Artificial Intelligence Powered Interfaces to Support Different Types of Memory for People with Dementia}

%%
%% The "author" command and its associated commands are used to define
%% the authors and their affiliations.
%% Of note is the shared affiliation of the first two authors, and the
%% "authornote" and "authornotemark" commands
%% used to denote shared contribution to the research.
\author{Hanuma Teja Maddali}
\email{hmaddali@umd.edu}
\affiliation{%
  \institution{University of Maryland}
  \city{College Park}
  \state{MD}
  \country{USA}
}

\author{Emma Dixon}
\email{eedixon@umd.edu}
\affiliation{%
  \institution{University of Maryland}
  \city{College Park}
  \state{MD}
  \country{USA}}
  
\author{Alisha Pradhan}
\email{alishapr@umd.edu}
\affiliation{%
  \institution{University of Maryland}
  \city{College Park}
  \state{MD}
  \country{USA}}
  
\author{Amanda Lazar}
\email{lazar@umd.edu}
\affiliation{%
  \institution{University of Maryland}
  \city{College Park}
  \state{MD}
  \country{USA}}

%%
%% By default, the full list of authors will be used in the page
%% headers. Often, this list is too long, and will overlap
%% other information printed in the page headers. This command allows
%% the author to define a more concise list
%% of authors' names for this purpose.
\renewcommand{\shortauthors}{Maddali, et al.}

%%
%% The abstract is a short summary of the work to be presented in the
%% article.
\begin{abstract}
There has been a growing interest in HCI to understand the specific technological needs of people with dementia and supporting them in self-managing daily activities. One of the most difficult challenges to address is supporting the fluctuating accessibility needs of people with dementia, which vary with the specific type of dementia and the progression of the condition. Researchers have identified auto-personalized interfaces, and more recently, Artificial Intelligence or AI-driven personalization as a potential solution to making commercial technology accessible in a scalable manner for users with fluctuating ability. However, there is a lack of understanding on the perceptions of people with dementia around AI as an aid to their everyday technology use and its role in their overall self-management systems, which include other non-AI technology, and human assistance. In this paper, we present future directions for the design of AI-based systems to personalize an interface for dementia-related changes in different types of memory, along with expectations for AI interactions with the user with dementia.
\end{abstract}

%%
%% The code below is generated by the tool at http://dl.acm.org/ccs.cfm.
%% Please copy and paste the code instead of the example below.
%%


%%
%% Keywords. The author(s) should pick words that accurately describe
%% the work being presented. Separate the keywords with commas.

%% A "teaser" image appears between the author and affiliation
%% information and the body of the document, and typically spans the
%% page.
%%\begin{teaserfigure}
%%  \includegraphics[width=\textwidth]{sampleteaser}
%%  \caption{Seattle Mariners at Spring Training, 2010.}
%%  \Description{Enjoying the baseball game from the third-base
%%  seats. Ichiro Suzuki preparing to bat.}
%%  \label{fig:teaser}
%%\end{teaserfigure}

%%
%% This command processes the author and affiliation and title
%% information and builds the first part of the formatted document.
\maketitle

\section{Introduction and Related Work}
The ability of people living with dementia to independently perform daily activities can vary depending on the sensory, motor, and cognitive changes specific to their type and stage of dementia \cite{who_dementia}. There has been growing interest among HCI researchers in understanding the specific needs of users with dementia when designing technology to support self-managing both activities of daily living and instrumental activities of daily living activities (ADLs and IADLs \cite{edemekong_activities_2022}) such as managing medication and online communication. Researchers have also begun to attend to the implications for designing personalized technology \cite{kerkhof_user-participatory_2019,oksnebjerg_assistive_2020, peterson_future_2012} to support meaningful activities for users with dementia, for example, social sharing \cite{lazar_supporting_2017} or leisure activities \cite{lazar_successful_2017}. A notable challenge is supporting the fluctuating accessibility needs of users with dementia when using dementia-specific assistive technology and, in general, commercially available devices such as smartphones and tablets \cite{dixon_role_2020}. The intensity with which a person experiences dementia-related changes in ability at any moment is not strictly predictable and can fluctuate with, for example, time of day, environmental stimuli, or mood. During these periods of fluctuating ability, the interface on a recently usable device can suddenly feel unusable for the user \cite{dixon_role_2020}. The impact on usability can be magnified for less tech-savvy users and can become a barrier to perform daily activities that require technology usage (e.g. searching for directions, video calling) \cite{archer_online_2014, guisado-fernandez_factors_2019, nygard_use_2007}. 
\begin{table*}
  \caption{Participant self-reported demographic information}
  \label{tab:demographics}
  \begin{tabular}{cclllll}
    \toprule
    ID & Age & Gender & Ethnicity & Country & Type of Dementia & Dementia Stage\\
    \midrule
    P1 & 63 & Male & Caucasian & UK & Mixed Vascular Dementia/Alzheimer's & Mild/Moderate\\
P2 & 65 & Female & Caucasian & UK & Alzheimer's & Mild/Moderate\\
P3 & 58 & Male & Caucasian & US & Lewy Body & Mild\\
P4 & 60 & Female & Caucasian & US & Subcortical Dementia & Unknown\\
P5 & 57 & Female & Caucasian & US & Younger Onset Alzheimer's & Mild\\
P6 & 59 & Female & Caucasian & US & Vascular Dementia/White Matter Disease & Mild/Moderate\\
P7 & 67 & Male & Caucasian & US & Vascular Dementia & Mild/Moderate\\
P8 & 67 & Female & Caucasian & US & Major Neuro-Cognitive Impairment & Mild/Moderate\\
P9 & 61 & Male & Caucasian & UK & Lewy Body & Mild\\
P10 & 61 & Male & Caucasian & US & Alzheimer's/Semantic Dementia & Mild\\
P11 & 59 & Male & Caucasian & US & Alzheimer's/Vascular Dementia & Moderate\\
P12 & 71 & Male & Caucasian & US & MCI \footnotemark[1] & Mild\\
P13 & 67 & Male & Caucasian & Canada & Vascular Cognitive & Moderate\\
P14 & 59 & Male & Caucasian & US & Early Onset Alzheimer's & Mild\\
P15 & 61 & Male & Caucasian & UK & Vascular Dementia & Mild\\
P16 & 73 & Female & Caucasian & US & Vascular Dementia & Mild\\
P17 & 55 & Female & African-American & US & MCI \footnotemark[2] & Mild\\
    \bottomrule
  \end{tabular}

\footnotemark[1] P12 has since been re-diagnosed as having Mild Cognitive Impairment (MCI). \footnotemark[2] P17 referred to herself as being in the early stages of dementia in the interview though she reported being diagnosed with MCI in the demographics form.
\end{table*}

Recent work has identified auto-personalization interfaces as a fruitful area for exploration to serve the fluctuating device accessibility needs of people with dementia and other disabilities \cite{dixon_role_2020}. For example, Morphic surfaces frequently used accessibility features for the Windows OS, such as existing Microsoft Word Immersive Reading tools, and allows users to save their preferred accessibility settings (e.g., text size and color contrast) to a hard-drive which automatically applies those settings to new machines \cite{vanderheiden_morphic_2020}. These are useful when the user is unfamiliar with accessibility features and needs to navigate layers of menus to find them, which Hu and Feng demonstrate can be more difficult for users with cognitive impairments \cite{hu_investigating_2015}. Researchers have also suggested auto-personalization, using Artificial Intelligence (AI), as a scalable approach to make mainstream technologies accessible rather than custom builds of dementia-specific assistive technologies \cite{dixon_taking_2021, vanderheiden_morphic_2020}. Dixon and Lazar \cite{dixon_role_2020} discuss the concept of an AI-driven interface that learns from the user with Dementia to improve interface usability. The AI-driven system would present content in a responsive manner to the user's in-situ device usability needs. For example, lowered ability to process words would prompt summarization of information sources. However, there is a lack of understanding on the perceptions of people with dementia around AI as an aid on their everyday devices and its role as part of the user's overall self-management approach that include other non-AI technology, and human assistance. 

This paper presents an analysis of transcripts from Dixon and Lazar's interview study on self-management of information needs by people with dementia \cite{dixon_taking_2021}. Prior HCI and sociology research has shown that multiple analysis through different perspectives albeit of the same data can yield different useful insights and allows wider use of data from rare or inaccessible participants such as people with dementia [12,16]. We explore Dixon and Lazar's study data from the angle of supporting different types of memory that might decline at different rates depending on the type of dementia and affect people's ability to perform different ADLs and IADLs in different ways \cite{giebel_activities_2015}.  There are two contributions of this paper related to the design of AI-driven interfaces for people living with dementia. First, we discuss how participants with early and moderate dementia perceive the role of different kinds of AI-driven technology when supporting activities related to different types of memory. This includes learnings from a conceptual prototype of an AI-driven auto-personalized interface presented to participants as part of a storyboard session. Second, we discuss future directions for the design of AI-driven personalized interfaces and environments based on expectations of people with dementia for these AI -driven interfaces .

\section{Study Details}
\subsection{Participants and Procedures}
We analyzed transcripts from Dixon and Lazar's study \cite{dixon_taking_2021} of sessions completed by 17 participants (average age = 62.5 years) with mild to moderate dementia (Table \ref{tab:demographics}) who self-reported being regular users of technology. Some participants identified as dementia advocates and even being members of advocacy organizations' peer-support groups in the past. All participants were screened to ensure their ability to give informed consent using the UC Davis Alzheimer's Disease Center procedures \cite{UCD_policy_2002}. 

\begin{figure*}[h]
\label{fig:storyboard}
  \centering
  \includegraphics[width=0.75\linewidth]{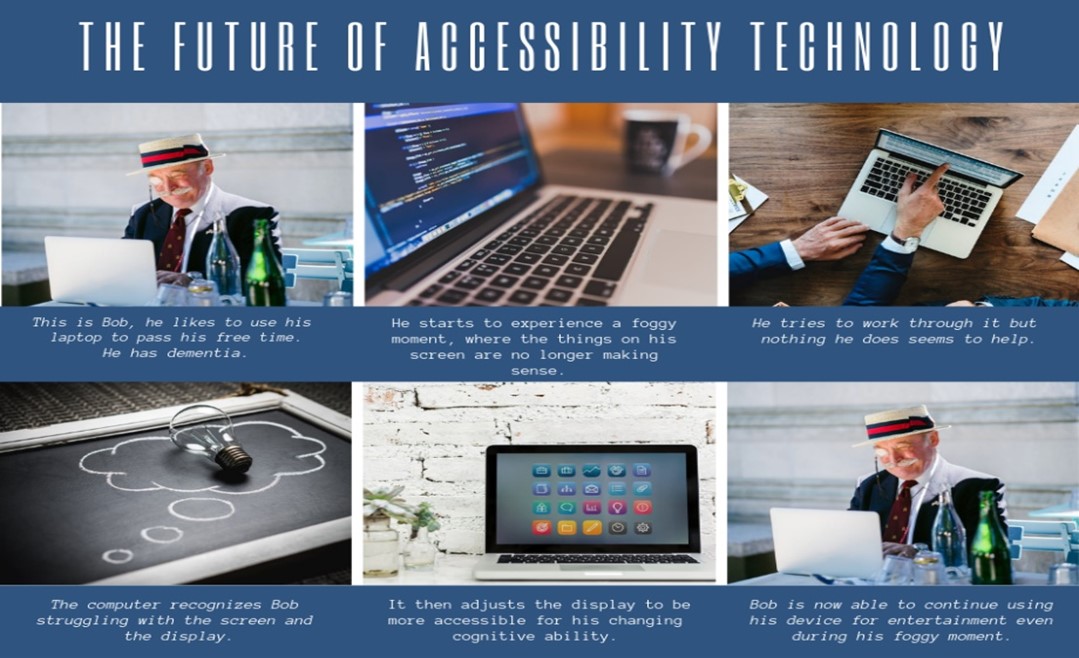}
  \caption{Storyboard illustrating AI that “simplifies” the interface when the user is experiencing a foggy moment.}
  \Description{The Figure shows a story board with six panels. Panel 1 has Bob, an older adult with dementia who is in front of a laptop that he uses to pass his free time. Panel 2 describes him experiencing a foggy moment where things on his screen no longer make sense, Panel 3 describes that he is trying to work through the confusion but nothing he does seems to help. Panel 4 visualizes an AI on the laptop that recognizes Bob struggling with the content on the interface. In Panel 5, the AI adjusts the display to be more accessible in some way for his changing cognitive ability. Panel 6 shows that Bob is now able to continue using his device for entertainment even during his foggy moment.}
\end{figure*}

Participants underwent a one-hour audio/video recorded semi-structured interview. Questions covered their approaches to self-managing daily activities, in addition to health and wellness, while living with dementia, the technologies they use for this purpose, and the limitations of existing technology for activities they might want to do. This was followed by the presentation of a storyboard (Figure \ref{fig:storyboard}) illustrating a conceptual prototype of an AI that detects moments of lowered ability and adapts the interface for improved accessibility. The storyboard was used to elicit feedback on the participants' perception of AI-driven accessible technologies. 

\subsection{Analysis}
A thematic analysis of the transcripts was performed with an initial inductive phase that included the first author creating open codes, memos, and themes and presenting them to the group of authors in an iterative manner. Two themes of “managing memory/reality” and “validating memory” emerged which seemed to categorize the data by different memory types. This was followed by a deductive phase of analysis where we use past literature on the influence of different memory types, prospective and retrospective memory, on daily activities for people with dementia \cite{nugent_home_2007} as a reference. This helped situate our discussion on perceptions around AI and its role for future accessible interfaces for user needs related to prospective and retrospective memory. 

\section{Findings}
We present findings on participants' perceptions about AI on devices they use every day and on the possibilities for future AI-driven auto-personalized interfaces for people with dementia as depicted by the conceptual prototype in Figure \ref{fig:storyboard}. First, we describe ideas around AI supporting two memory types: prospective and retrospective memory. These memory types appear in various other past works exploring the connection between ADLs, IADLs, and different types of memory \cite{giebel_activities_2015, nugent_home_2007, van_der_roest_assistive_2017}. It needs to be noted here that voice-based AI assistants (e.g. Amazon Alexa, Google home) figure prominently as the representative examples for the form and capabilities of AI driven interfaces that participants are most familiar with. Second, we describe expectations for AI-assistance when compared with human assistance.
\subsection{Supporting Prospective Memory by Reminding and Planning with AI}
\textbf{Prospective memory} allows us to make plans, retain them, and bring them back to one's consciousness in response to the right cues \cite{uttl_prospective_2018}. This can be essential to tasks that are habitual (e.g., brushing teeth, medication), require vigilance (e.g., preventing bathtub overflow), and need to be performed at a specific time or place (e.g., at a grocery store, in the afternoon). The idea that an AI could gather information about the user's daily routine and habits and converse with the user to remind them when required was often brought up in participants' discussions. Many participants imagined the presence of AI in their homes as a positive when living with dementia. P3 talked about an AI that could prompt him with a \textit{“"Hey, [P3], don't forget to take your medication this morning". I mean it has to be seamless. It has to be happening without me even realizing it's happening”}. P6 talks about \textit{“a grid of normal, day to day tasks that you were, would automatically be reminded of ...You know if somehow something would ping, “What are you having for dinner tonight? Going out? Staying in?”} Participants also thought about how an AI could monitor and alert the users to tasks that were left unfinished like \textit{“Hey, did you close the refrigerator door? Right. Did you turn that stove off? … Oh, I didn't hear I didn't hear the snap on the thing [fridge door or stove top turning off]. You know”} (P10).  In addition to the type of cue, when and where a type of cue is presented is something that an AI needed to consider. Participants talk about their existing ways of leaving notes or having others leave reminders where they are likely to see them. Having these cues or reminders appear audibly or in a visibly at the right time and location is an important consideration. For example, P14 thought it would be helpful \textit{“if there was something that would ding as I'm walking out, or when I approached to do something that would help us and it would actually say, [P14], don't forget, you know, be at Lowe's. And 3:30, you have a doctor (appointment) .”} P9 even mentions adding custom prompts himself to his Alexa AI assistant to inject \textit{“a sense of humanness”} and say something like \textit{“[P9], you need to have a shower or you will smell”}. 

The cost of prospective memory lapses can be high especially in social interactions when living with dementia. Planning for social events could involve, for example, using a digital calendar to schedule events, and remembering context that includes who will be attending and what might be expected of attendees. P4 even talks about how this prep is important for doing something like dressing the part for the event as she couldn't read these social cues as well as before. AI was also seen as something that would help the user keep engaged socially, for example, by reminding them to call friends they haven't talked with in a while by saying something \textit{“We've noticed that you haven't spoke with [Friend] in three weeks or so. May I suggest a call?”} (P6). In addition to the AI interacting with the user with dementia, we also find participants having thoughts about how its role in their social settings and how it should interact with others beside the primary user. Some participants preferred using an AI assistant like Alexa, something they already owned, that approached users through conversation, for example, when looking for instructions for a task. This doesn't break the flow of their activity or require them to remember something to search for online before returning to the task. The level of visibility required of the AI as a social actor and how often it injects itself into a situation depends on the use. As P15 explains, it could be the friendly, reassuring voice of a \textit{“gentle companion”} that isn't too intrusive but instead disappears into the background \textit{“when you're actually physically trying to whatcha call it, trying to concentrate on something”}. Devices that are used by both people with dementia and care partners together (e.g., an AI assistant like Google Home, or Amazon Alexa) can act as an intermediary voice for the care partner as P11 explains: \textit{“She [his wife] went in unbeknownst to me, and she put a message on there that says, at noon, “eat [P11] eat” because if it wasn't there, I would not eat lunch.”} 

\subsection{Supporting Retrospective Memory }
\textbf{Retrospective memory} is the memory of people, words, and events encountered or experienced in the past \cite{uttl_prospective_2018}. It covers the ability to encode and recollect long-term past events and working memory to remember information for short duration tasks. It also includes semantic memory related to facts, ideas, meaning and concepts (e.g., recalling words or numbers to understand language). An important note is that retrospective and prospective memories are not considered to be entirely independent. There is a retrospective component to prospective memory as well because recalling a plan is essential to be able to respond to it. Participants describe verifying thoughts (e.g. long-term memories) and adapting to changes in their semantic and working memory as part of perspectives on using AI to support retrospective memory. 
\subsubsection{Verifying thoughts}
The inability to recall memories of life events or process words (e.g., names) at the right moment affected participants' social interactions and sense of identity. In the different, social and technological approaches that participants used to support retrospective memory, we find that participants highlighted the importance of preserving emotional context, specifically for long-term memory. This was something \textit{“inherently present when remembering something on your own”} (P9) but might not be adequately captured by an audio or video recording. Participants also talk about instances when other people associated with a memory, helped them remember the emotional context required to comprehend what they were experiencing when they couldn't recall this context by themselves. For example, P11's sister helped provide this emotional context when P11 talks about a time when he couldn't understand why he became unbelievably emotional and started crying at Disney world. It was a place that held a lot of memories for him and where he could \textit{“hear the music that you grew up on”} and \textit{“smell those familiar odors”}. It took a phone call to his sister \textit{“to remind me of why that place had that effect on me”}. This was an interesting example of how a (remote) loved one was helping recall and providing comfort in an emotionally overwhelming situation. P2 also talks about how \textit{"if I saw a newspaper article that somebody sent me by my phone and I would phone them up and I would say, "Tell me more, tell me more!". Especially when I've forgotten I'd even been there of even done it."} Even when something like a photo doesn't bring back memory that's gone, rather than getting upset P2 explains that it was helpful for acceptance that a certain experience did happen \textit{“because I've got evidence [photo]”}. Participants talk about conversing with their voice assistant AI in a similar manner to help with moments of confusion around their own identity, memories, and state of mind. P2 mentioned that she might ask her google assistant \textit{"Hey Google, what's my name?"} or \textit{"Hey Google, where do I live?"} if she is struggling to remember. P9 even talks about using his Alexa voice assistant to create a journal that acts as \textit{“virtual memory”} and can include something as mundane as his partner's name or play an entry from a particular day to \textit{“experience a repetition of emotion and that's when the memory becomes a memory for me and is no longer just a piece of info.”}  We also find other interesting instances of using technology to assist with “plugging the gaps” (P1). This can be, for example, when there's some word (e.g. name, place, object) they can't recall. In which case they can, for example, search the internet or ask an AI assistant a question about it. P2 even imagined a scenario where an AI assistant detects a confused user trying to recall the name of a drink and presenting multiple answers like \textit{“coffee, tea, squash ...”} that they could choose from. On a more day-to-day basis P4 explains how a combination of tools helps to triangulate and verify information scanned in the moment: \textit{“So, I might scan for the time I think I'm supposed to go some place and I think I've got it. But, I've got it wrong. But, when Alexa says it, I hear it, and I can see it in my calendar for example to double check it.”} 
\subsubsection{Adapting to changes in semantic and working memory on a visible interface}
More frequently we encounter examples of participants temporarily losing their capacity to comprehend, remember, or communicate information using the right words. Participants mention that it can happen when communicating with a doctor (P7) or  when you are out at a restaurant and \textit{“you having trouble reading the menu” }(P11). Short-term or semantic memory constraints also affected how participants prepared for conversation and meetings. P3, for example, tries to prepare for these situations by looking at the menu ahead of time \textit{“because a lot of menus are not dementia friendly”}. We find similar examples that could directly impact device usability when participants discussed various possible nuances related to AI-driven interface \textit{“simplification”} as presented in the storyboard (Figure \ref{fig:storyboard}). For example, verbal or text instructions on a device might not be understandable at a certain moment. Participant feedback on the conceptual prototype presented in the storyboard included more specific strategies for an AI that transformed the interface content into more meaningful personalized combinations of text, icons, or audio. For some participants, meaningful icons would be easier to understand over text when dealing with a mental fog. P2 for instance explains that in such a state, words can sometimes be meaningless and \textit{“if I was experiencing this sort of trouble, if I saw a lightbulb or a cloud [as in the first panel on the second row of the storyboard in Figure \ref{fig:storyboard}] I would just like to sort of screen touch it and then the pictures would help me enormously.”}  When reminding the user of an activity, P16 suggested a similar idea of being able to see \textit{“a picture that shows somebody brushing their teeth or taking their pills or whatever, like on a daily calendar.”} P4 on the other hand says directing the user with words \textit{"touch here"} or \textit{"press here to choose this option"} might be more useful and if the interface tried to oversimplify images, they might just look that same. At the end of the day, the cues used by the AI \textit{“are useless to someone who doesn't know what they are”} (P4) and need to be meaningful enough to the intended user. P3 for example, suggests that a smart home could detect when he was agitated and do something more tailored to him such as \textit{“automatically turn on my television and show a video of, a prerecorded video of my wife talking to me that would help calm me down. Or even better, it could send a text message to [wife's name] and she could call me.”}  For some, it was important for the AI to preemptively check with the user before intervening. For example, notifying the user before it decides to call for help or “simplifies” a device interface to adapt to the constraints of the user's working or semantic memory. So, conversation or intentional cuing between the user and AI was preferred so the user with dementia can expect and prepare for a change in the interface. P7 explains that \textit{“mind wise, it preps me to get ready to change into another mode.”} 

Participants gave other examples related to constraints on their short-term/working memory and describe that the AI could also simplify the interface in terms or provide a smaller set of options for menus. Participants seemed to prefer technology that, in general, required less steps to remember when operating. An example of this was their need for simpler control over privacy settings that might force the users, like P16, to \textit{“go through 100 or whatever choices on setting privacy and all of this, when it could just be? I only want to see and be contacted by the pages that I want to contact? Why couldn't that be the first choice.”} Relying on trial and error to understand what each choice on an interface results in can be cumbersome as P10 points out: \textit{“you make a mistake because you hit the wrong thing. You remember to go back and hit the right thing the next time. Well, I could do that four or five times … hit that same wrong thing each time because I just don't recall what I just did.”} The nuance in simplifying is in the AI understanding the user's personal preferences or whatever seems to make more sense based on the user's ability at that time. It is important that an AI simplifying an interface doesn't do so in a way that seems patronizing, like for a child and is designed using the same principles as for an adult: \textit{“don't make it so that you think the person is simple but simplify the process (for the user)”} (P15).  

\subsection{Expectations for AI compared with human assistance}
When comparing technological assistance (AI or no-AI) with human assistance, participants talk about different expectations. Expectations for AI were related to how much agency people are willing to give up, or an estimation of how an AI can augment their abilities. While technology does allow participants to feel less dependent on human assistance, it is also necessary and many times better to involve the caregiver in some situations that require empathetic vigilance. P11 talks about how during moments of \textit{“very thick fog”}, a caregiver can sometimes sense the gravity of the situation better since \textit{“all she has to do is look at me and she knows she knows what's going on and she springs in action.”} The context for some activities can also be better suited for a loved one to help with any memory needs without feeling dependent on someone else. P3 describes cooking together with this wife as \textit{“a blast, we love it”}, and where \textit{“she helps, because you know there have been plenty of instances where too much of a certain ingredient goes in or it goes in twice, or you know things like that.”} 

When it came to AI, we find contrasting thoughts on whether the idea of making it more “human” in some sense was useful. The storyboard (Figure \ref{fig:storyboard}) presented the concept of an AI that can recognize when the user needs assistance which some participants understood as a situation where AI that was sensitive to anxiety (P3), inactivity (P4, P6), confusion (P2). While participants like P13 think that the AI's interpretation of what its sensing \textit{“can lead to misunderstandings”}, an equally legitimate point brought up by P9 was that \textit{“people always assume they know best for you. Alexa doesn't make any assumption like that at the moment.”} Interactions with AI can lack an escalation of confrontation so \textit{“a person living with Alzheimer's can take out their Alzheimer's anger on a non-human.”} A care partner, who consented to be part of the study, spoke up during P11's session on how this missing expectation of confrontation can make it \textit{“easier for [P11] sometimes to be in his own world with an electronic device than it is to interact with me or it's not just me, it's me human.”} She stresses the need for a balance in human and non-human interactions that isn't detrimental to the user's interactions with their loved ones and other human beings. Finally, trusting an AI and the \textit{“algorithm behind the AI”}, can be a bit more ambiguous with some participants being more cynical about sharing their data. Others like P3 felt \textit{"the benefits far outweigh the risk"} and were even excited by the idea of an AI in the background that could seamlessly assist the user when observing them: \textit{“I can think of even the webcam being used to know when to read a person's face to know when they're frustrated. to know when they're confused.”}

\section{Discussion and Conclusion}
Our findings raise interesting future directions for AI-driven personalization of dementia-accessible interfaces and environments. Given that participants in Dixon and Lazar's study were using voice-based AI assistants to support prospective memory-based tasks, our findings suggest there are further opportunities to customize the context of this AI assistance. Researchers have begun to investigate AI-based contextual reminders for people with dementia, such as Carroll et al.'s conceptual designs for the Robin system \cite{carroll_robin_2017}, which provides verbal assistance (e.g. relay step-by-step instruction) for routine tasks or Donaldson's system \cite{donaldson_assistive_2018} using a home sensor network to recognize and complete incomplete tasks (e.g. turning off the lights) automatically. Our work provides justification to expand efforts in this space, designing systems that combine verbal assistance for routine tasks with home sensor networks to provide meaningful cues at the right time and location to effectively support the prospective memory of people with dementia. Further, our findings inform future work in this area by describing the perspectives of people with dementia on the tasks they are comfortable with receiving AI-assistance, ways to provide more meaningful cues, and their comfort with different approaches to AI interpreting their need for assistance. These are all important considerations when designing future AI-driven adaptive interfaces. 

Another consideration for future AI-driven adaptive interfaces is learning user preferences for the mode of information presentation related to changes in retrospective semantic and working memory. This is of particular importance when designing AI-driven adaptive interfaces for people with dementia, as researchers in this space have demonstrated the interplay of sensory and cognitive changes people with dementia experience affecting accessible modes of information presentation \cite{dixon_role_2020}. Our findings provide examples of different potential approaches AI could take to unobtrusively learn about the user's preferences for the mode of information presentation: 1) automatically and seamlessly collecting information through sensors in a smart home interfaced using a voice-based AI assistant, 2) conversing with the user directly and asking about their preferences for accessible interfaces (e.g. icons over words, simplifying explorable choices), and 3) interacting with friends or family members who might also use the same device for their own needs (e.g. adding events or custom reminders). There is a relatively stronger representation of the approach of automatic collection of data using sensor networks in literature \cite{donaldson_assistive_2018, meiland_technologies_2017, nugent_home_2007}. However, we join with recent work \cite{dixon_taking_2021} in calling designers and developers to explore the other two approaches that center the needs and wishes of individuals with dementia on how AI learns about their information presentation needs in order to support individuals privacy preferences and their need to be self-determinate.   

A final consideration for future AI assistant interactions which emerged from the data was the need for “humanness” of the AI assistant in its interactions with the user. This becomes especially important in user's expectations that an AI be aware of the social cost of memory lapses, on perceptions of how well the AI can interpret the user's state to offer assistance, and in how it presented itself (e.g., as a friendly gentle voice). A related point of discussion are how user preferences for visibility of the AI, in terms of its pro-activeness and as a social actor can interact with the sense of agency of a user with dementia when being assisted \cite{coghlan_dignity_2021}. For future research, this can mean exploring, for example, when an AI could automatically adapt an interface versus scenarios when notifying the user and obtaining permission before doing so is desirable. Often AI assistants like Amazon Alexa or Google assistant are designed with certain human like personalities which play a role in how users perceive interacting with them \cite{pradhan_hey_2021}. There is an opportunity for researchers to explore how such personalities as well as voice of these technologies can be designed to better support people with dementia. In this space, there has been past work proposing automated systems that converse with the users with dementia to support retrospective memory (e.g. reminiscence \cite{bermingham_automatically_2013, caros_automatic_2020}). Malhotra et. al's findings \cite{malhotra_exploratory_2015} also call to understand how emotionally aligned prompts can be designed and delivered when assisting step-wise with daily activities (e.g. hand washing) through virtual-human assistants. Based on our findings, we suggest that future research utilize participatory design methods to better understand user-desired AI personalities (e.g., having a partner's voice)  to better provide the necessary emotional context when supporting users in completing tasks related to retrospective memory.

%%
%% The acknowledgments section is defined using the "acks" environment
%% (and NOT an unnumbered section). This ensures the proper
%% identification of the section in the article metadata, and the
%% consistent spelling of the heading.
\begin{acks}
This work was supported, in part, by grant 90REGE0008, U.S. Admin. for Community Living, NIDILRR, Dept. of Health and Human Services, and National Science Foundation Grants IIS-2045679 and DGE-1840340. Opinions expressed do not necessarily represent official policy of the Federal government.
\end{acks}

%%
%% The next two lines define the bibliography style to be used, and
%% the bibliography file.
\bibliographystyle{ACM-Reference-Format}
\bibliography{proceedings}

%%
%% If your work has an appendix, this is the place to put it.
%%\appendix

%%\section{Research Methods}

%%\subsection{Part One}

%%Lorem ipsum dolor sit amet, consectetur adipiscing elit. Morbi
%%malesuada, quam in pulvinar varius, metus nunc fermentum urna, id
%%sollicitudin purus odio sit amet enim. Aliquam ullamcorper eu ipsum
%%vel mollis. Curabitur quis dictum nisl. Phasellus vel semper risus, et
%%lacinia dolor. Integer ultricies commodo sem nec semper.

\end{document}